\documentclass[aps,pra,twocolumn,showpacs,amsmath,amssymb,preprintnumbers,superscriptaddress,10pt]{revtex4-1}

\usepackage{amsmath,amssymb}
\usepackage{bm}
\usepackage[latin1]{inputenc}
\usepackage{graphicx}
\graphicspath{{figures/}}
\usepackage{SIunits}
\usepackage{hyphenat}
\usepackage[bookmarks=false]{hyperref}
\usepackage{color}
\usepackage[capitalise]{cleveref}
\crefname{section}{Sec.}{Secs.}
\Crefname{section}{Section}{Sections}

\usepackage{pifont} 
\usepackage[nomessages]{fp} 
\newcommand{\circled}[1]{\FPadd\p{181}{#1}\FPupn\p{\p{} 0 round}\raisebox{-0.3ex}{{\Large \ding{\p}}}}
\newcommand{\circledincaption}[1]{\FPadd\p{181}{#1}\FPupn\p{\p{} 0 round}\raisebox{-0.27ex}{{\large \ding{\p}}}} 

\definecolor{pink}{RGB}{255,0,255}

\definecolor{ss_color}{rgb}{1,0,0}

\definecolor{ngreen}{rgb}{0.2,0.6,0.2}

\definecolor{ngreen2}{rgb}{0.2,0.6,0.6}

\begin{document}

\title{Security loophole in free-space quantum key distribution due to\\
spatial-mode detector-efficiency mismatch}

\author{Shihan~Sajeed}
\email{ssajeed@uwaterloo.ca}
\affiliation{Institute for Quantum Computing, University of Waterloo, Waterloo, ON, N2L~3G1 Canada}
\affiliation{\mbox{Department of Electrical and Computer Engineering, University of Waterloo, Waterloo, ON, N2L~3G1 Canada}}

\author{Poompong~Chaiwongkhot}
\affiliation{Institute for Quantum Computing, University of Waterloo, Waterloo, ON, N2L~3G1 Canada}
\affiliation{Department of Physics and Astronomy, University of Waterloo, Waterloo, ON, N2L~3G1 Canada}

\author{Jean-Philippe~Bourgoin}
\affiliation{Institute for Quantum Computing, University of Waterloo, Waterloo, ON, N2L~3G1 Canada}
\affiliation{Department of Physics and Astronomy, University of Waterloo, Waterloo, ON, N2L~3G1 Canada}

\author{Thomas~Jennewein}
\affiliation{Institute for Quantum Computing, University of Waterloo, Waterloo, ON, N2L~3G1 Canada}
\affiliation{Department of Physics and Astronomy, University of Waterloo, Waterloo, ON, N2L~3G1 Canada}
\affiliation{Quantum Information Science Program, Canadian Institute for Advanced Research, Toronto, ON, M5G~1Z8 Canada}

\author{Norbert~L{\" u}tkenhaus}
\affiliation{Institute for Quantum Computing, University of Waterloo, Waterloo, ON, N2L~3G1 Canada}
\affiliation{Department of Physics and Astronomy, University of Waterloo, Waterloo, ON, N2L~3G1 Canada}

\author{Vadim~Makarov}
\affiliation{Institute for Quantum Computing, University of Waterloo, Waterloo, ON, N2L~3G1 Canada}
\affiliation{Department of Physics and Astronomy, University of Waterloo, Waterloo, ON, N2L~3G1 Canada}
\affiliation{\mbox{Department of Electrical and Computer Engineering, University of Waterloo, Waterloo, ON, N2L~3G1 Canada}}

\date{\today}

\begin{abstract}
In free-space quantum key distribution (QKD), the sensitivity of the receiver's detector channels may depend differently on the spatial mode of incoming photons. Consequently, an attacker can control the spatial mode to break security. We experimentally investigate a standard polarization QKD receiver, and identify sources of efficiency mismatch in its optical scheme. We model a practical intercept-and-resend attack and show that it would break security in most situations. We show experimentally that adding an appropriately chosen spatial filter at the receiver's entrance may be an effective countermeasure.
\end{abstract}

\maketitle

\section{Introduction}
\label{sec:introduction}

Quantum key distribution (QKD) \cite{bennett1984,ekert1991}, in theory, allows two distant parties Alice and Bob to establish a shared secret key with unconditional security \cite{lo1999, shor2000, lutkenhaus2000, mayers2001, renner2005}. Although a number of successful implementations of QKD have been reported \cite{bennett1992b,gobby2004,schmitt-manderbach2007,stucki2009} and commercialization is underway \cite{comqkdsystems2015}, the technology is yet to achieve widespread use. One important reason is that the maximum distance is still of the order of $300~\kilo\meter$ \cite{shibata2014} in fiber-based systems. Consequently, implementation of free-space QKD utilizing ground-to-satellite links \cite{buttler1998,kurtsiefer2002,kurtsiefer2002a,hughes2002,weier2006,ursin2007,erven2008,peloso2009,nauerth2013} that promises long-distance quantum communication is now a very attractive field of research.

Implementation imperfections have enabled a number of successful attacks on QKD \cite{sajeed2015,jain2014,jouguet2013,sun2011,lydersen2010a,zhao2008,qi2007,makarov2006,vakhitov2001,*gisin2006}. The main reason behind this is the deviation of the actual behaviour of the devices from the ideal expected behaviour. Thus, to guarantee the security, it is of utmost importance to scrutinize the practical device behaviours for possible deviations. One such source of deviation in free-space QKD can be the assumed symmetry of detection efficiency among all received quantum states in Bob's detector \cite{makarov2006,lamas-linares2007,qi2007,zhao2008,weier2011}. If a deviation from this assumption exists, an adversary Eve can send light to Bob in different spatial modes so that one of his detectors has a relatively higher probability of click than the other detector(s) \cite{fung2009}. In this way, she can exploit the mismatch in efficiency and make Bob's measurement outcome dependent on his measurement basis and correlated to Eve, which breaks the assumptions of typical security proofs. In this work, we investigate how crucial this can be to the security of QKD. (While finishing this paper, we became aware of a recent similar work \cite{rau2015}.)

We study a receiver designed for polarization encoding free-space QKD, described in \cref{sec:qkd_system}. We test it in \cref{sec:experiment} by sending an attenuated laser beam to the receiver with various angle offsets and recording the relative detection probability in each channel, to find incidence angles with high efficiency mismatch. With these data, we show in \cref{sec:modeling} by numerical modeling that an eavesdropper attack exists that enables Eve to steal the secret key. We discuss countermeasures in \cref{sec:countermeasures} and conclude in \cref{sec:conclusion}.

\section{QKD system under test}
\label{sec:qkd_system}

\begin{figure*}
\centering
\includegraphics{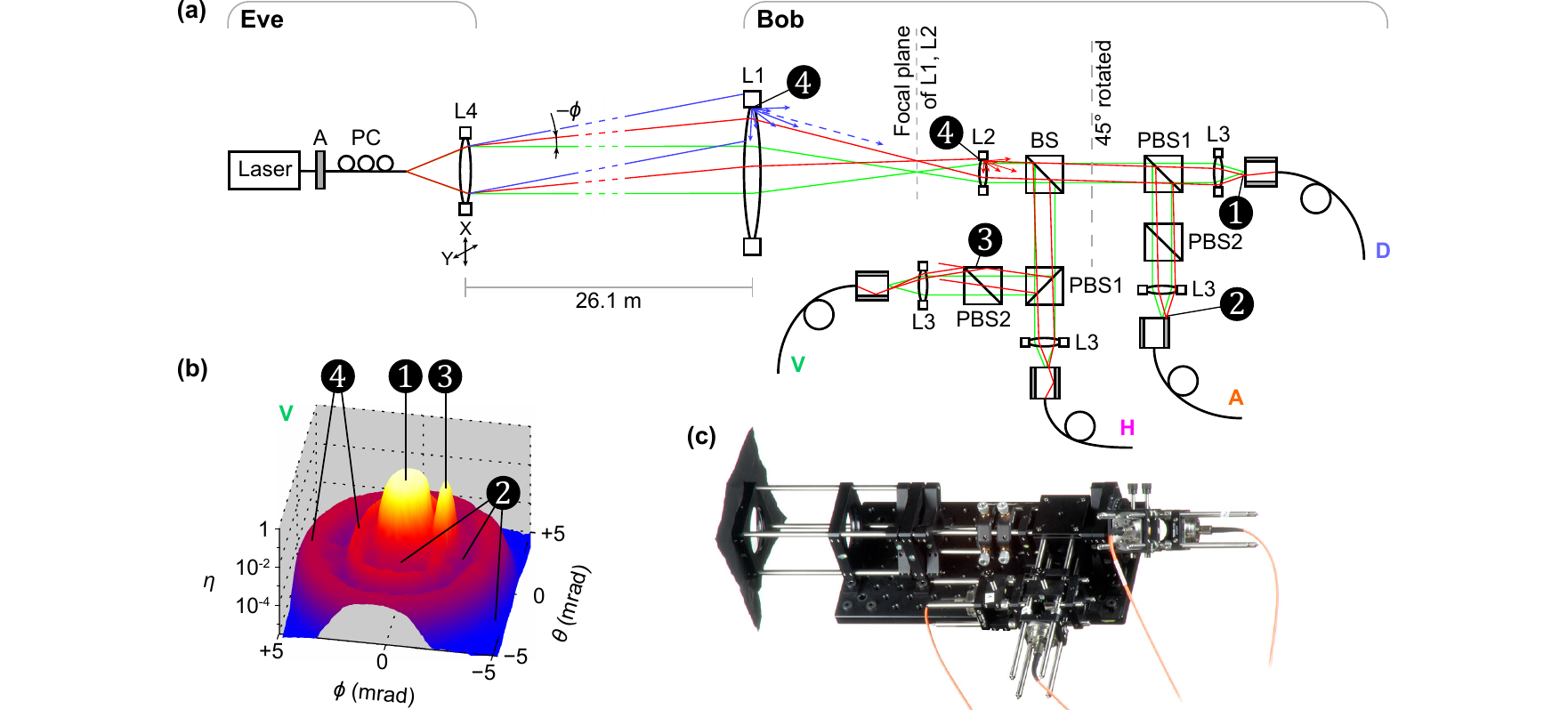}
\caption{(color online). Experimental setup. (a)~Scheme of the experimental apparatus, top view (drawing not to scale). Eve's source consists of a fiber-coupled $532~\nano\meter$ laser, attenuator A, polarization controller PC, and a collimating lens mounted on a two-axis motorised translation stage. The latter allows changing the beam's incidence angle and lateral displacement at Bob's front lens L1 simultaneously. Green (light gray) marginal rays parallel to the optical axis denote the original alignment of Alice's beam to Bob. Red and blue (dark gray) marginal rays show a scanning beam from Eve tilted at an angle $(\phi, \theta)$ relative to the original beam. Features \protect\circledincaption{1}--\protect\circledincaption{4} mark different transmission paths for light inside Bob. (b)~Normalized detection efficiency $\eta$ in channel V versus the illumination angle $(\phi, \theta)$. This scan was taken to show the features clearly by placing Eve at a closer distance. (c)~Photograph of Bob's receiver. The actual distance between facing surfaces of L2--BS is $42~\milli\meter$, BS--PBS1 $66~\milli\meter$, PBS1--L3 $31~\milli\meter$, PBS1--PBS2 $45~\milli\meter$, PBS2--L3 $10~\milli\meter$ in channel A and $5~\milli\meter$ in channel V.} 
\label{fig:setup}
\end{figure*}

A free-space QKD receiver typically employs a telescope to reduce the size of a collimated beam, followed by a non-polarizing beamsplitter to randomly choose between two measurement bases. It is followed by polarization beamsplitters and single-photon detectors to measure photons in the four states of polarization: horizontal (H), vertical (V), $+45\degree$ (D), and $-45\degree$ (A) \cite{buttler1998,kurtsiefer2002,kurtsiefer2002a,hughes2002,weier2006,ursin2007,erven2008,peloso2009,nauerth2013}. The receiver we test is a prototype for a quantum communication satellite \cite{bourgoin2014}, operating at $532~\nano\meter$ wavelength [\cref{fig:setup}(a,c)]. Its telescope consists of a focusing lens L1 (diameter $50~\milli\meter$, focal length $f = 250~\milli\meter$; Thorlabs AC508-250-A) and collimating lens L2 ($f = 11~\milli\meter$; Thorlabs A397TM-A). The collimated beam of $\lesssim 2~\milli\meter$ diameter then passes through a 50:50 beamsplitter BS (custom pentaprism \cite{bourgoin2014}) and pairs of polarization beamsplitters PBS1 and PBS2 (Thorlabs PBS121). PBS2 increases the polarization extinction ratio in the reflected arm of PBS1. Lenses L3 (Thorlabs PAF-X-18-PC-A) focus the four beams into $105~\micro\meter$ core diameter multimode fibers (Thorlabs M43L01) leading to single-photon detectors (Excelitas SPCM-AQRH-12-FC).

Long-distance free-space QKD receivers are multimode for two reasons. First, propagation of Alice's beam, initially single-mode, through a turbulent atmosphere splits it into multiple spatial modes \cite{tyson2010}. Second, the finite precision and speed of real-time angular tracking of Alice's beam requires that Bob accepts multiple spatial modes in a certain acceptance angle \cite{bienfang2004,weier2006,ursin2007,nauerth2013}. Use of single-mode fibers under these conditions would lead to additional coupling losses $\gtrsim 10~\deci\bel$ \cite{takenaka2012} if the system does not include appropriate (and often expensive) adaptive correction optics \cite{tyson2010}. Therefore, multimode fibers and detectors with larger area are generally preferred as they allow good collection efficiency without increasing complexity and cost.

\section{Experiment}
\label{sec:experiment}

In order to exploit the mismatch in efficiency, Eve needs to know the mismatch for the four detectors as a function of the input angle. Hence, our first step was to scan Bob's receiver for possible efficiency mismatch. Eve's source [\cref{fig:setup}(a)] consists of a $532~\nano\meter$ laser coupled into single-mode fiber, attenuator A, polarization controller PC, and a collimating lens L4 (Thorlabs C220TME-A) mounted on a two-axis motorised translation stage (Thorlabs MAX343/M). In \cref{fig:setup}(a), green (light gray) marginal rays denote the initial alignment from Eve, replicating the alignment from Alice to Bob. This is the initial position of the translation stage $\phi = \theta = 0$. As we moved the stage in the transverse plane, it changed the beam's incidence angle and lateral displacement at Bob's front lens L1 simultaneously. This is shown by red (dark gray) marginal rays in \cref{fig:setup}(a), representing a beam from Eve coming at an angle $(\phi, \theta)$ relative to the initial beam. 

\begin{figure*}
\centering
\includegraphics{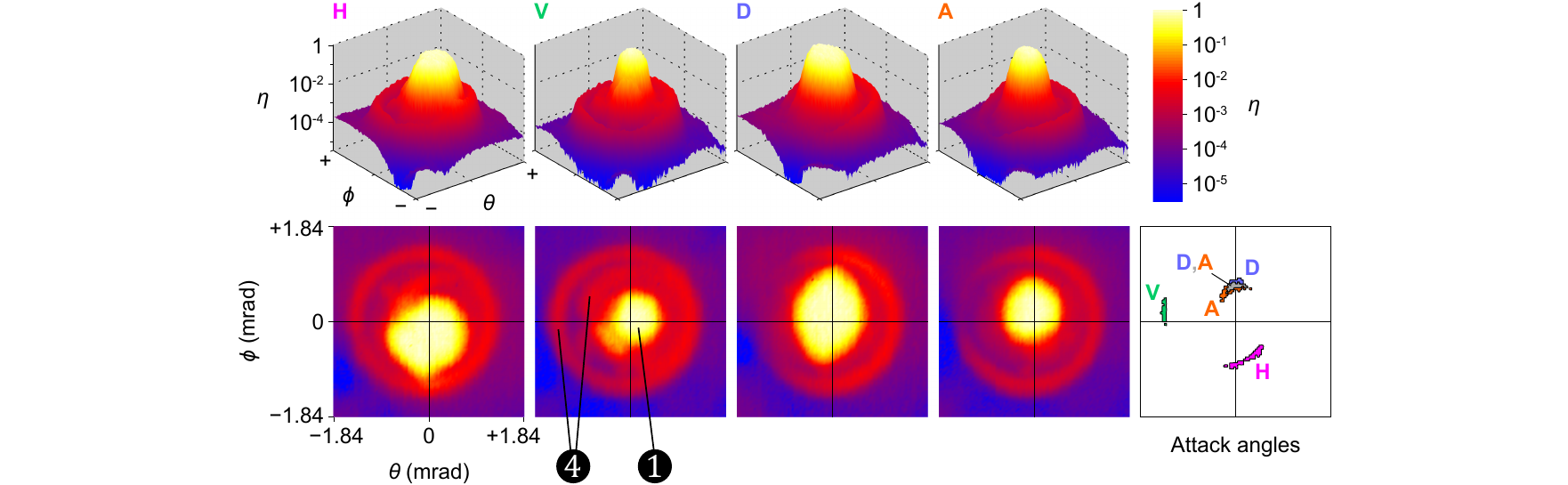}
\caption{(color online). Angular efficiency scan of the receiver, and points of interest. Four pair of plots {\bf H, V, D, A} shown in both 3D and 2D represent normalized detection efficiency in the four receiver channels versus illuminating beam angle $(\phi, \theta)$. The angle $\phi = \theta = 0$ is the initial angle of QKD operation. The last plot shows angle ranges with a high mismatch, usable in our attack.}
\label{fig:scanning}
\end{figure*}

Before scanning, the optics in Bob's apparatus was aligned to maximize coupling into all four detectors at the normal incidence, which is the standard alignment procedure for QKD. Note that many free-space QKD systems employ a real-time tracking system to maintain this initial alignment \cite{bienfang2004,weier2006,ursin2007,nauerth2013}. We then started the scanning procedure that involved first, changing the outgoing beam's angle $(\phi, \theta)$, and then recording the corresponding count rate at all four detectors of Bob. For each data point, we used an integration time of $1~\second$. Our scan consisted of approximately $100 \times 100$ data points in a square matrix covering the whole clear aperture of Bob's front lens L1. Then during post-processing, for each data point for each detector, we subtracted the corresponding detector's background count rate, and then normalized it by dividing by the maximum count rate in that detector.

At first, we did a preliminary scan using optical power meters (Thorlabs PM200 with S130C head) that revealed several features, highlighted in \cref{fig:setup}(b). Around $\phi = \theta = 0$, maximum light coupling resulted in the central peak \circled{1}. With increasing scanning angle, the focused beam started missing the fiber core, and the detector count dropped off \circled{2}. A region was found when the beam reflected off a polished edge of PBS2 back into the fiber core, causing the peak \circled{3}. Increasing the angle further made the beam hit the anodized aluminum mount of L1 and possibly edges of other lens mounts and round elements in the optical assembly. It was scattered at these edges, producing two ring-like features \circled{4}. Beyond these features, there were no noticeable power reading, as the beam completely missed the receiver aperture.

We then adjusted the receiver setup to minimize the peak \circled{3}, and performed final scans at $26.1~\meter$ distance using Bob's single-photon detectors (Excelitas SPCM-AQRH-12-FC). During these scans, the beam at L1 was Gaussian-shaped with $9~\milli\meter$ width (at $1/e^2$ peak intensity). The scans were done in $38.3~\micro\radian$ steps covering $\pm 1.84~\milli\radian$ range, corresponding to lateral displacement of $\pm 48~\milli\meter$ at L1. \Cref{fig:scanning} shows the normalized detection efficiency in all four receiver channels as a function of $(\phi, \theta)$. Most of the original features are still visible. However, outside the narrow central range of angles close to $\phi = \theta = 0$, individual channel's efficiencies vary independently. Also, the size and shape of the central peak is significantly different between channels. This was impossible to identify during the normal alignment procedure. This effect can be attributed to imprecise focusing, optical path length difference between the arms, off-centered alignment of lenses, mode-dependent bending loss in fibers, and individual variations in components. These may have also caused the efficiency at one side of the outer ring being higher. Because of these reasons, there exist angles such that if photons are sent at those angles, one channel has a much higher click probability than the rest.

\section{Attack model}
\label{sec:modeling}

To emphasize the security threat, it is useful to model an attack that exploits the discovered side-channel. One possible attack is the faked-state attack \cite{makarov2006, makarov2005}, which is an intercept-and-resend attack in which Eve attempts to deterministically control Bob's basis choice and detection outcome. We model a practical faked-state attack using the obtained data and the following assumptions: Alice and Bob perform non-decoy-state Bennett-Brassard 1984 (BB84) protocol using polarization encoding. Alice emits weak coherent pulses with mean photon number $\mu$ equal to Alice--Bob line transmittance \cite{lutkenhaus2000}. Whenever Bob registers a multiple click, he performs a squashing operation (double-click in one basis is mapped to a random value in that basis, while multiple clicks in different bases are discarded) \cite{beaudry2008,tsurumaru2008,gittsovich2014}. Alice and Bob also monitor total sifted key rate, and quantum bit error ratio (QBER). Eve has information about Bob's receiver characteristics described above, and only uses devices available in today's technology. She intercepts photons at the output of Alice, using an active basis choice and superconducting nanowire detectors, with overall detection efficiency $\eta_e = 0.85$ and dark count probability $< 10^{-9}$ per bit slot \cite{marsili2013.NatPhotonics-7-210}. Then, a part of her, situated close to Bob, regenerates the measured signal and sends to Bob. We assume that Alice--Bob and Alice--Eve fidelity $F = 0.9831$ \cite{bourgoin2014}, while Eve--Bob experimentally measured $F = 0.9904$. Here fidelity refers to the probability that a polarized photon will emerge from the PBS at the correct path, which is related to visibility by $F = (1+\text{visibility})/2$. We also confirmed experimentally that Eve--Bob fidelity is preserved at all illumination angles shown in \cref{fig:scanning}.

From Eve's point of view, she wants to maximize the detection probability when Bob measures in compatible (i.e.,\ same as her) basis, to maximize Eve--Bob mutual information. Also, she wants to minimize Bob's detection probability in non-compatible basis, to minimize QBER. Let $\eta_i(j)$ be the efficiency of Bob's $i$-th channel ($i \in \{h,v,d,a\}$) given that incoming light is $j \in \{H,V,D,A\}$ polarized. Thus to find attack points for the $j$-th polarization, we choose angles that have higher values of $\eta_j(j)$ and $\delta_j(j) = \min \left\{ \frac{\eta_j(j)}{\eta_{nc0}(j)}, \frac{\eta_j(j)}{\eta_{nc1}(j)} \right\}$, where $\eta_{nc0}$ and $\eta_{nc1}$ are the normalized efficiencies of the two detectors in the non-compatible basis. Our experimental attack angles are shown in the rightmost plot in \cref{fig:scanning}. For example, the H attack angles were composed of points for which $\eta_h(H) \geq 0.2$ and $\delta_h(H) \geq 75$. Similarly, for the V, D and A attack angles, $\eta_v(V) \geq 0.002$, $\delta_V \geq 8$; $\eta_d(D) \geq 0.4$, $\delta_D \geq 80$; $\eta_a(A) \geq 0.1$, $\delta_A \geq 20$. The thresholds used here to find the attack angles were not optimal, and were picked manually. 

To derive the key rate and QBER formula in Eve's presence, we start with a system with only Eve and Bob. Let's consider Eve sending an $H$-polarized pulse to Bob within the attack angles H. Before squashing, the raw click probability $p_i(j)$ that detector $i$ in Bob clicks given Eve has sent $j$-polarized light is
\begin{equation}
\begin{aligned}
&p_h(H) \approx c_h+ 1 - \exp\left(-\frac{\mu_H F \eta_h(H)}{2}\right),\\
&p_v(H) \approx c_v+1 - \exp\left(-\frac{\mu_H (1-F) \eta_v(H)}{2}\right),\\
&p_{d(a)}(H) \approx c_{d(a)}+1 - \exp\left(-\frac{\mu_H \eta_{d(a)}(H)}{4}\right),
\end{aligned}
\end{equation}
where $\mu_H$ is Eve's mean photon number and $c_i$ is Bob's background click probability per bit slot in $i$-th channel. The probability $P_{hv}(H)$ that after squashing Bob measures in HV basis, given Eve has sent an $H$-polarized pulse, is composed of three events: when only detector H clicks, when only detector V clicks, or when both click. It can be written as
\begin{equation}
\begin{aligned}
P_{hv}(H) &= \big[ 1-p_d(H) \big] \big[ 1-p_a(H) \big]\\
     &\quad\, \times \big[ p_h(H)+p_v(H) - p_h(H) p_v(H) \big].
\end{aligned}
\end{equation}

Let's now include Alice into the picture. Consider Alice sends an $H$-polarized pulse, and Eve intercepts it. Let $P_c^e \approx \frac{1}{2} (1-e^{-\mu F \eta_e }) e^{-\mu (1-F) \eta_e}$ and $P_w^e \approx \frac{1}{2} e^{-\mu F \eta_e} (1-e^{-\mu (1-F) \eta_e})$ be the probability that Eve measures in the compatible basis (i.e.,\ the same basis as Alice) and gets a click only in the correct and wrong detector respectively. Let $P_{nc}^e \approx \frac{1}{2} (1-e^{ -\frac{\mu \eta_e}{2}}) e^{-\frac{\mu \eta_e}{2}}$ be the probability that she measures in the non-compatible basis (different basis than Alice's) and gets a click in a single detector. The sifted key rate given Alice has sent $H$-polarized light is
\begin{equation} \label{eq:rate_h}
\begin{aligned}
\! R_e(H) \approx \, & P_c^e P_{hv}(H) + P_w^e P_{hv}(V) + P_{nc}^e \left[ P_{hv}(D)\!+\!P_{hv}(A)\right]\\
&  + (1- P_c^e - P_w^e - 2 P_{nc}^e) (c_h + c_v - c_h c_v).
\end{aligned}
\end{equation}
An error can occur when Eve measures Alice's signal in non-compatible basis or when Eve measures in compatible basis but Bob measures a wrong value owing to imperfect fidelity or dark count. Hence, the error rate conditioned on Alice sending $H$-polarized light is
\begin{equation} \label{eq:error_h}
\begin{aligned}
E_H \approx \, & P_c^e P_v(H) + P_w^e P_v(V) + P_{nc}^e \left[ P_v(D) + P_v(A)\right] \\
& + (1- P_c^e - P_w^e - 2 P_{nc}^e) (c_v - \frac{c_v c_h}{2}),
\end{aligned}
\end{equation}
where $P_i(j)$ is the probability that Bob measures value $i$ after squashing, given Eve has sent $j$-polarized light. For example,
\begin{equation}
P_v(H) = \big[ p_v(H) - \frac{p_h(H) p_v(H)}{2} \big] \big[ 1-p_d(H) \big] \big[ 1-p_a(H) \big].
\end{equation}
Sifted key rates and errors in Eve's presence [\cref{eq:rate_h,eq:error_h}] conditioned on $V$, $D$, $A$ polarizations sent by Alice can be calculated similarly. The total sifted key rate and QBER in Eve's presence become
\begin{equation}
\begin{aligned}
&R_e = \frac{1}{4} \sum_{j = H,V,D,A} R_e(j),\\
&\text{QBER}_e =\frac{1}{4 R_e} \sum_{j = H,V,D,A} E_j.
\end{aligned}
\end{equation}

\begin{figure}
\centering
\includegraphics{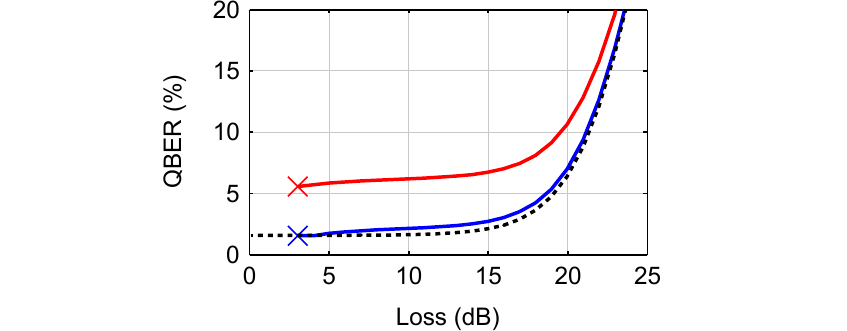}
\caption{(color online). Modeled QBER observed by Bob versus line loss. The dotted curve shows QBER without Eve. At lower line loss, the QBER is due to imperfect fidelity, while at higher line loss Bob's detector background counts become the dominant contribution. The lower solid curve (blue) shows $\text{QBER}_e$ under our attack when only the total Bob's sifted key rate $R_{ab}$ is matched. The upper solid curve (red) additionally keeps his four channel rates equal.}
\label{fig:loss_error}
\end{figure}

\begin{figure*}
\centering
\includegraphics{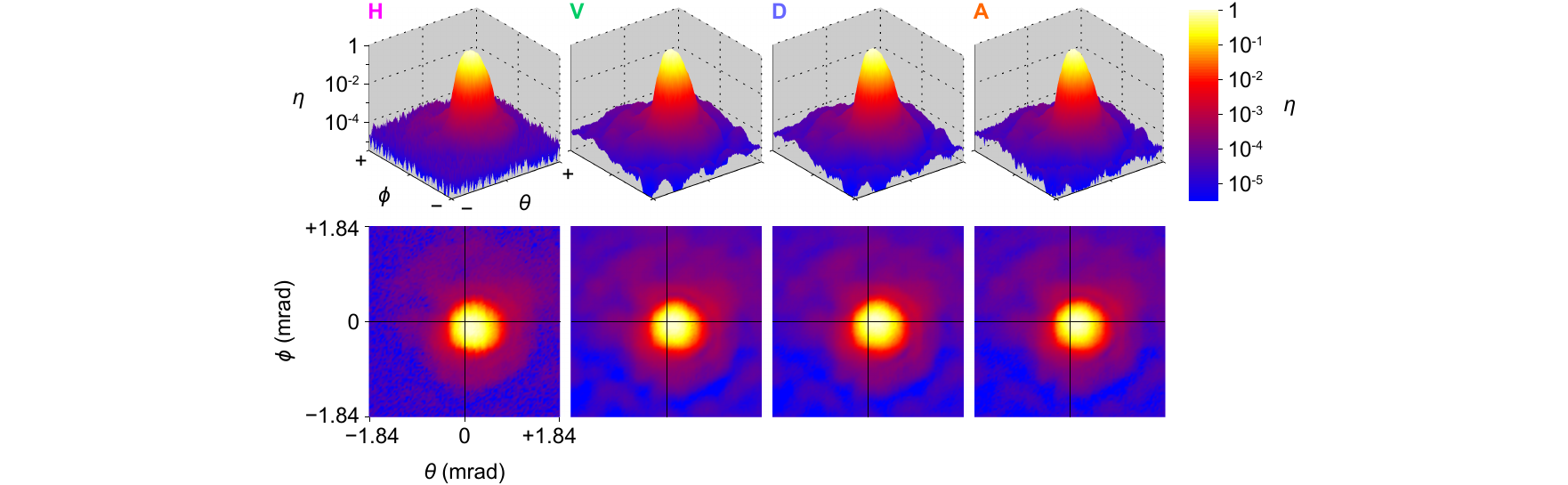}
\caption{(color online). Angular efficiency scan of the receiver after a $25~\micro\meter$ diameter pinhole (Thorlabs P25S) is placed in the focal plane of L1, L2 [\cref{fig:setup}(a)]. No detectable mismatch between channels was found under tight search conditions $\eta_i(j) \geq 0.001$ and $\delta_i(j) \geq 4$.}
\label{fig:pinhole}
\end{figure*}

The only free parameters left for Eve to manipulate are the mean photon numbers of her signal. Knowing the angular scanning data, Eve can use a numerical optimization to find values of $\mu_H$, $\mu_V$, $\mu_D$, $\mu_A$ that minimize $\text{QBER}_e$ while keeping $R_e = R_{ab}$, where $R_{ab}$ is Bob's sifted key rate without Eve. Our numerical optimization achieves this for Alice--Bob channel loss $\geq 3~\deci\bel$ if they are willing to accept a slight increase of QBER by less than $0.7\%$ (see \cref{fig:loss_error}). Here we assumed Bob's detector parameters as measured by us: efficiency at $\phi = \theta = 0$ was $0.4$ in all four channels, and individual detector background count probabilities were in the range of $430 \times 10^{-9}$ to $1560 \times 10^{-9}$ per $1~\nano\second$ coincidence window. These optimization results are realistic conditions for a successful attack on most communication channels \cite{buttler1998,kurtsiefer2002,kurtsiefer2002a,hughes2002,ursin2007,erven2008,nauerth2013,bourgoin2014} Note that the distance Eve--Bob can be increased without affecting attack performance, by replacing Eve's illuminator with four collimators oriented at the required attack angles. 

We went further and imposed an additional constraint on Eve to make $R_e(H) = R_e(V) = R_e(D) = R_e(A) = R_{ab}$. Our optimization shows that it is still possible for Eve to pick appropriate mean photon numbers and successfully attack the system with resultant QBER $< 6.82\%$ in $3$--$15~\deci\bel$ line loss range (\cref{fig:loss_error}). Similar QBER values are typical for outdoor channels, because of background light. Eve could shield Bob from the latter to hide QBER resulting from her attack.

We would like to point out that the attack angles depend on the way the setup is constructed, the imperfections of each individual sample of component, and each individual alignment procedure. I.e.,\ no two setups are identical, even if they are produced in the same assembly line, and they will generally have different attack angles. However, from a theoretical point of view, in quantum cryptography it is assumed from Kerckhoffs' principle \cite{kerckhoffs1883} that except for the keys themselves, Eve has knowledge about all other parameters in the system. It is thus a valid assumption that she knows the attack angles. From a practical point of view, Eve may try techniques proposed in \cite{makarov2005}. She may replace a small fraction of the signal states with faked states at different spatial angles, then listen to the classical communication to get an estimate of the efficiency of Bob's detectors at those angles. In this way she may gradually improve her estimate on the mismatch without causing excessive QBER. When she has enough information on the statistics of the mismatch, she can launch her full-fledged attack.

\section{Countermeasures}
\label{sec:countermeasures}

In our attack, by sending lights at different angles, Eve has broken a fundamental assumption of security proofs that detection probabilities are independent of detection basis \cite{tomamichel2012,inamori2007}. We propose to restore this assumption by placing a spatial filter (pinhole) at the focal plane of Bob's L1 and L2 [\cref{fig:setup}(a)]. Spatial filtering is sometimes done before the beamsplitters to increase signal-to-background ratio in the channel \cite{hughes2002,weier2006,peloso2009}, however it has not been characterised as a security countermeasure. We performed scanning with $100$, $75$, and $25~\micro\meter$ diameter pinholes, and found that decreasing the pinhole diameter gradually reduces the mismatch. The $25~\micro\meter$ diameter pinhole eliminated any visible mismatch (\cref{fig:pinhole}) even though we reduced our search parameters to $\eta_i(j) \geq 0.001$ and $\delta_i \geq 4$. This pinhole provides Bob's field-of-view of $100~\micro\radian$, which does not reduce his efficiency with turbulent atmospheric channels \cite{ursin2007}. Hence, we conclude that a $25~\micro\meter$ pinhole may be an efficient countermeasure for the current setup. 

Note that, in Refs.~\onlinecite{qi2007} and \onlinecite{dasilva2015}, a detector scrambling strategy was proposed that might be an effective countermeasure against efficiency mismatch attacks for single-photon qubits. However, it is not clear how effective that countermeasure is, when one considers that the detectors operate on optical modes, not on single-photon signals. This can be a future study.

\section{Conclusion}
\label{sec:conclusion}

Our analysis implies that data obtained during a QKD session can be explained by an intercept-resend attack exploiting the spatial mode side-channels. Therefore, there is no postprocessing or privacy amplification that can eliminate Eve's knowledge without sacrificing all key \cite{curty2004a}. Although our practical attack should work, and the physical countermeasure seems promising, there is still room for improvement on both the attack scheme and countermeasures. Eve can employ more attack angles or combine this attack with some other suitable attack schemes, to increase the number of her free parameters. Alice and Bob can make this harder by monitoring more parameters. We expect that our attack can be conducted also in the related decoy-state protocol \cite{hwang2003}, though the requirement to match the correct decoy statistics will modify the parameter regime where it will be effective. Another possible future study is to fully implement the present attack under realistic outdoor channel conditions.

\begin{acknowledgments}
We thank Y.~Zhang and M.~Mosca for discussions. This work was supported by US Office of Naval Research, Industry Canada, CFI, Ontario MRI, NSERC, Canadian Space Agency, and CryptoWorks21. P.C.\ acknowledges support by Thai DPST scholarship. J.-P.B.\ and T.J.\ acknowledge support from FED~DEV.
\end{acknowledgments}


\begin{thebibliography}{51}%
\makeatletter
\providecommand \@ifxundefined [1]{%
 \@ifx{#1\undefined}
}%
\providecommand \@ifnum [1]{%
 \ifnum #1\expandafter \@firstoftwo
 \else \expandafter \@secondoftwo
 \fi
}%
\providecommand \@ifx [1]{%
 \ifx #1\expandafter \@firstoftwo
 \else \expandafter \@secondoftwo
 \fi
}%
\providecommand \natexlab [1]{#1}%
\providecommand \enquote  [1]{``#1''}%
\providecommand \bibnamefont  [1]{#1}%
\providecommand \bibfnamefont [1]{#1}%
\providecommand \citenamefont [1]{#1}%
\providecommand \href@noop [0]{\@secondoftwo}%
\providecommand \href [0]{\begingroup \@sanitize@url \@href}%
\providecommand \@href[1]{\@@startlink{#1}\@@href}%
\providecommand \@@href[1]{\endgroup#1\@@endlink}%
\providecommand \@sanitize@url [0]{\catcode `\\12\catcode `\$12\catcode
  `\&12\catcode `\#12\catcode `\^12\catcode `\_12\catcode `\%12\relax}%
\providecommand \@@startlink[1]{}%
\providecommand \@@endlink[0]{}%
\providecommand \url  [0]{\begingroup\@sanitize@url \@url }%
\providecommand \@url [1]{\endgroup\@href {#1}{\urlprefix }}%
\providecommand \urlprefix  [0]{URL }%
\providecommand \Eprint [0]{\href }%
\providecommand \doibase [0]{http://dx.doi.org/}%
\providecommand \selectlanguage [0]{\@gobble}%
\providecommand \bibinfo  [0]{\@secondoftwo}%
\providecommand \bibfield  [0]{\@secondoftwo}%
\providecommand \translation [1]{[#1]}%
\providecommand \BibitemOpen [0]{}%
\providecommand \bibitemStop [0]{}%
\providecommand \bibitemNoStop [0]{.\EOS\space}%
\providecommand \EOS [0]{\spacefactor3000\relax}%
\providecommand \BibitemShut  [1]{\csname bibitem#1\endcsname}%
\let\auto@bib@innerbib\@empty
\bibitem [{\citenamefont {Bennett}\ and\ \citenamefont
  {Brassard}(1984)}]{bennett1984}%
  \BibitemOpen
  \bibfield  {author} {\bibinfo {author} {\bibfnamefont {C.~H.}\ \bibnamefont
  {Bennett}}\ and\ \bibinfo {author} {\bibfnamefont {G.}~\bibnamefont
  {Brassard}},\ }in\ \href@noop {} {\emph {\bibinfo {booktitle} {Proceedings of
  IEEE International Conference on Computers, Systems, and Signal
  Processing}}}\ (\bibinfo  {publisher} {IEEE Press, New York},\ \bibinfo
  {address} {Bangalore, India},\ \bibinfo {year} {1984})\ pp.\ \bibinfo {pages}
  {175--179}\BibitemShut {NoStop}%
\bibitem [{\citenamefont {Ekert}(1991)}]{ekert1991}%
  \BibitemOpen
  \bibfield  {author} {\bibinfo {author} {\bibfnamefont {A.~K.}\ \bibnamefont
  {Ekert}},\ }\href {\doibase 10.1103/PhysRevLett.67.661} {\bibfield  {journal}
  {\bibinfo  {journal} {Phys. Rev. Lett.}\ }\textbf {\bibinfo {volume} {67}},\
  \bibinfo {pages} {661} (\bibinfo {year} {1991})}\BibitemShut {NoStop}%
\bibitem [{\citenamefont {Lo}\ and\ \citenamefont {Chau}(1999)}]{lo1999}%
  \BibitemOpen
  \bibfield  {author} {\bibinfo {author} {\bibfnamefont {H.-K.}\ \bibnamefont
  {Lo}}\ and\ \bibinfo {author} {\bibfnamefont {H.~F.}\ \bibnamefont {Chau}},\
  }\href {\doibase 10.1126/science.283.5410.2050} {\bibfield  {journal}
  {\bibinfo  {journal} {Science}\ }\textbf {\bibinfo {volume} {283}},\ \bibinfo
  {pages} {2050} (\bibinfo {year} {1999})}\BibitemShut {NoStop}%
\bibitem [{\citenamefont {Shor}\ and\ \citenamefont
  {Preskill}(2000)}]{shor2000}%
  \BibitemOpen
  \bibfield  {author} {\bibinfo {author} {\bibfnamefont {P.~W.}\ \bibnamefont
  {Shor}}\ and\ \bibinfo {author} {\bibfnamefont {J.}~\bibnamefont
  {Preskill}},\ }\href {\doibase 10.1103/PhysRevLett.85.441} {\bibfield
  {journal} {\bibinfo  {journal} {Phys. Rev. Lett.}\ }\textbf {\bibinfo
  {volume} {85}},\ \bibinfo {pages} {441} (\bibinfo {year} {2000})}\BibitemShut
  {NoStop}%
\bibitem [{\citenamefont {L\"utkenhaus}(2000)}]{lutkenhaus2000}%
  \BibitemOpen
  \bibfield  {author} {\bibinfo {author} {\bibfnamefont {N.}~\bibnamefont
  {L\"utkenhaus}},\ }\href {\doibase 10.1103/PhysRevA.61.052304} {\bibfield
  {journal} {\bibinfo  {journal} {Phys. Rev. A}\ }\textbf {\bibinfo {volume}
  {61}},\ \bibinfo {pages} {052304} (\bibinfo {year} {2000})}\BibitemShut
  {NoStop}%
\bibitem [{\citenamefont {Mayers}(2001)}]{mayers2001}%
  \BibitemOpen
  \bibfield  {author} {\bibinfo {author} {\bibfnamefont {D.}~\bibnamefont
  {Mayers}},\ }\href {\doibase 10.1145/382780.382781} {\bibfield  {journal}
  {\bibinfo  {journal} {J. ACM}\ }\textbf {\bibinfo {volume} {48}},\ \bibinfo
  {pages} {351} (\bibinfo {year} {2001})}\BibitemShut {NoStop}%
\bibitem [{\citenamefont {Renner}\ \emph {et~al.}(2005)\citenamefont {Renner},
  \citenamefont {Gisin},\ and\ \citenamefont {Kraus}}]{renner2005}%
  \BibitemOpen
  \bibfield  {author} {\bibinfo {author} {\bibfnamefont {R.}~\bibnamefont
  {Renner}}, \bibinfo {author} {\bibfnamefont {N.}~\bibnamefont {Gisin}}, \
  and\ \bibinfo {author} {\bibfnamefont {B.}~\bibnamefont {Kraus}},\ }\href
  {\doibase 10.1103/PhysRevA.72.012332} {\bibfield  {journal} {\bibinfo
  {journal} {Phys. Rev. A}\ }\textbf {\bibinfo {volume} {72}},\ \bibinfo
  {pages} {012332} (\bibinfo {year} {2005})}\BibitemShut {NoStop}%
\bibitem [{\citenamefont {Bennett}\ \emph {et~al.}(1992)\citenamefont
  {Bennett}, \citenamefont {Bessette}, \citenamefont {Salvail}, \citenamefont
  {Brassard},\ and\ \citenamefont {Smolin}}]{bennett1992b}%
  \BibitemOpen
  \bibfield  {author} {\bibinfo {author} {\bibfnamefont {C.~H.}\ \bibnamefont
  {Bennett}}, \bibinfo {author} {\bibfnamefont {F.}~\bibnamefont {Bessette}},
  \bibinfo {author} {\bibfnamefont {L.}~\bibnamefont {Salvail}}, \bibinfo
  {author} {\bibfnamefont {G.}~\bibnamefont {Brassard}}, \ and\ \bibinfo
  {author} {\bibfnamefont {J.}~\bibnamefont {Smolin}},\ }\href {\doibase
  10.1007/bf00191318} {\bibfield  {journal} {\bibinfo  {journal} {J.
  Cryptology}\ }\textbf {\bibinfo {volume} {5}},\ \bibinfo {pages} {3}
  (\bibinfo {year} {1992})}\BibitemShut {NoStop}%
\bibitem [{\citenamefont {Gobby}\ \emph {et~al.}(2004)\citenamefont {Gobby},
  \citenamefont {Yuan},\ and\ \citenamefont {Shields}}]{gobby2004}%
  \BibitemOpen
  \bibfield  {author} {\bibinfo {author} {\bibfnamefont {C.}~\bibnamefont
  {Gobby}}, \bibinfo {author} {\bibfnamefont {Z.~L.}\ \bibnamefont {Yuan}}, \
  and\ \bibinfo {author} {\bibfnamefont {A.~J.}\ \bibnamefont {Shields}},\
  }\href {\doibase 10.1063/1.1738173} {\bibfield  {journal} {\bibinfo
  {journal} {Appl. Phys. Lett.}\ }\textbf {\bibinfo {volume} {84}},\ \bibinfo
  {pages} {3762} (\bibinfo {year} {2004})}\BibitemShut {NoStop}%
\bibitem [{\citenamefont {Schmitt-Manderbach}\ \emph
  {et~al.}(2007)\citenamefont {Schmitt-Manderbach}, \citenamefont {Weier},
  \citenamefont {F\"urst}, \citenamefont {Ursin}, \citenamefont {Tiefenbacher},
  \citenamefont {Scheidl}, \citenamefont {Perdigues}, \citenamefont {Sodnik},
  \citenamefont {Kurtsiefer}, \citenamefont {Rarity}, \citenamefont
  {Zeilinger},\ and\ \citenamefont {Weinfurter}}]{schmitt-manderbach2007}%
  \BibitemOpen
  \bibfield  {author} {\bibinfo {author} {\bibfnamefont {T.}~\bibnamefont
  {Schmitt-Manderbach}}, \bibinfo {author} {\bibfnamefont {H.}~\bibnamefont
  {Weier}}, \bibinfo {author} {\bibfnamefont {M.}~\bibnamefont {F\"urst}},
  \bibinfo {author} {\bibfnamefont {R.}~\bibnamefont {Ursin}}, \bibinfo
  {author} {\bibfnamefont {F.}~\bibnamefont {Tiefenbacher}}, \bibinfo {author}
  {\bibfnamefont {T.}~\bibnamefont {Scheidl}}, \bibinfo {author} {\bibfnamefont
  {J.}~\bibnamefont {Perdigues}}, \bibinfo {author} {\bibfnamefont
  {Z.}~\bibnamefont {Sodnik}}, \bibinfo {author} {\bibfnamefont
  {C.}~\bibnamefont {Kurtsiefer}}, \bibinfo {author} {\bibfnamefont {J.~G.}\
  \bibnamefont {Rarity}}, \bibinfo {author} {\bibfnamefont {A.}~\bibnamefont
  {Zeilinger}}, \ and\ \bibinfo {author} {\bibfnamefont {H.}~\bibnamefont
  {Weinfurter}},\ }\href {\doibase 10.1103/PhysRevLett.98.010504} {\bibfield
  {journal} {\bibinfo  {journal} {Phys. Rev. Lett.}\ }\textbf {\bibinfo
  {volume} {98}},\ \bibinfo {pages} {010504} (\bibinfo {year}
  {2007})}\BibitemShut {NoStop}%
\bibitem [{\citenamefont {Stucki}\ \emph {et~al.}(2009)\citenamefont {Stucki},
  \citenamefont {Walenta}, \citenamefont {Vannel}, \citenamefont {Thew},
  \citenamefont {Gisin}, \citenamefont {Zbinden}, \citenamefont {Gray},
  \citenamefont {Towery},\ and\ \citenamefont {Ten}}]{stucki2009}%
  \BibitemOpen
  \bibfield  {author} {\bibinfo {author} {\bibfnamefont {D.}~\bibnamefont
  {Stucki}}, \bibinfo {author} {\bibfnamefont {N.}~\bibnamefont {Walenta}},
  \bibinfo {author} {\bibfnamefont {F.}~\bibnamefont {Vannel}}, \bibinfo
  {author} {\bibfnamefont {R.~T.}\ \bibnamefont {Thew}}, \bibinfo {author}
  {\bibfnamefont {N.}~\bibnamefont {Gisin}}, \bibinfo {author} {\bibfnamefont
  {H.}~\bibnamefont {Zbinden}}, \bibinfo {author} {\bibfnamefont
  {S.}~\bibnamefont {Gray}}, \bibinfo {author} {\bibfnamefont {C.~R.}\
  \bibnamefont {Towery}}, \ and\ \bibinfo {author} {\bibfnamefont
  {S.}~\bibnamefont {Ten}},\ }\href {\doibase 10.1088/1367-2630/11/7/075003}
  {\bibfield  {journal} {\bibinfo  {journal} {New J. Phys.}\ }\textbf {\bibinfo
  {volume} {11}},\ \bibinfo {pages} {075003} (\bibinfo {year}
  {2009})}\BibitemShut {NoStop}%
\bibitem [{com()}]{comqkdsystems2015}%
  \BibitemOpen
  \href@noop {} {}\bibinfo {note} {Commercial QKD systems are available for
  purchase, as of 2015, from at least three entities: ID~Quantique
  (Switzerland), \url{http://www.idquantique.com}; SeQureNet (France),
  \url{http://www.sequrenet.com}; and Austrian Institute of Technology
  (Austria), \url{http://www.ait.ac.at/}.}\BibitemShut {Stop}%
\bibitem [{\citenamefont {Shibata}\ \emph {et~al.}(2014)\citenamefont
  {Shibata}, \citenamefont {Honjo},\ and\ \citenamefont
  {Shimizu}}]{shibata2014}%
  \BibitemOpen
  \bibfield  {author} {\bibinfo {author} {\bibfnamefont {H.}~\bibnamefont
  {Shibata}}, \bibinfo {author} {\bibfnamefont {T.}~\bibnamefont {Honjo}}, \
  and\ \bibinfo {author} {\bibfnamefont {K.}~\bibnamefont {Shimizu}},\ }\href
  {\doibase 10.1364/OL.39.005078} {\bibfield  {journal} {\bibinfo  {journal}
  {Opt. Lett.}\ }\textbf {\bibinfo {volume} {39}},\ \bibinfo {pages} {5078}
  (\bibinfo {year} {2014})}\BibitemShut {NoStop}%
\bibitem [{\citenamefont {Buttler}\ \emph {et~al.}(1998)\citenamefont
  {Buttler}, \citenamefont {Hughes}, \citenamefont {Kwiat}, \citenamefont
  {Lamoreaux}, \citenamefont {Luther}, \citenamefont {Morgan}, \citenamefont
  {Nordholt}, \citenamefont {Peterson},\ and\ \citenamefont
  {Simmons}}]{buttler1998}%
  \BibitemOpen
  \bibfield  {author} {\bibinfo {author} {\bibfnamefont {W.~T.}\ \bibnamefont
  {Buttler}}, \bibinfo {author} {\bibfnamefont {R.~J.}\ \bibnamefont {Hughes}},
  \bibinfo {author} {\bibfnamefont {P.~G.}\ \bibnamefont {Kwiat}}, \bibinfo
  {author} {\bibfnamefont {S.~K.}\ \bibnamefont {Lamoreaux}}, \bibinfo {author}
  {\bibfnamefont {G.~G.}\ \bibnamefont {Luther}}, \bibinfo {author}
  {\bibfnamefont {G.~L.}\ \bibnamefont {Morgan}}, \bibinfo {author}
  {\bibfnamefont {J.~E.}\ \bibnamefont {Nordholt}}, \bibinfo {author}
  {\bibfnamefont {C.~G.}\ \bibnamefont {Peterson}}, \ and\ \bibinfo {author}
  {\bibfnamefont {C.~M.}\ \bibnamefont {Simmons}},\ }\href {\doibase
  10.1103/PhysRevLett.81.3283} {\bibfield  {journal} {\bibinfo  {journal}
  {Phys. Rev. Lett.}\ }\textbf {\bibinfo {volume} {81}},\ \bibinfo {pages}
  {3283} (\bibinfo {year} {1998})}\BibitemShut {NoStop}%
\bibitem [{\citenamefont {Kurtsiefer}\ \emph
  {et~al.}(2002{\natexlab{a}})\citenamefont {Kurtsiefer}, \citenamefont
  {Zarda}, \citenamefont {Halder}, \citenamefont {Weinfurter}, \citenamefont
  {Gorman}, \citenamefont {Tapster},\ and\ \citenamefont
  {Rarity}}]{kurtsiefer2002}%
  \BibitemOpen
  \bibfield  {author} {\bibinfo {author} {\bibfnamefont {C.}~\bibnamefont
  {Kurtsiefer}}, \bibinfo {author} {\bibfnamefont {P.}~\bibnamefont {Zarda}},
  \bibinfo {author} {\bibfnamefont {M.}~\bibnamefont {Halder}}, \bibinfo
  {author} {\bibfnamefont {H.}~\bibnamefont {Weinfurter}}, \bibinfo {author}
  {\bibfnamefont {P.~M.}\ \bibnamefont {Gorman}}, \bibinfo {author}
  {\bibfnamefont {P.~R.}\ \bibnamefont {Tapster}}, \ and\ \bibinfo {author}
  {\bibfnamefont {J.~G.}\ \bibnamefont {Rarity}},\ }\href {\doibase
  10.1038/419450a} {\bibfield  {journal} {\bibinfo  {journal} {Nature}\
  }\textbf {\bibinfo {volume} {419}},\ \bibinfo {pages} {450} (\bibinfo {year}
  {2002}{\natexlab{a}})}\BibitemShut {NoStop}%
\bibitem [{\citenamefont {Kurtsiefer}\ \emph
  {et~al.}(2002{\natexlab{b}})\citenamefont {Kurtsiefer}, \citenamefont
  {Zarda}, \citenamefont {Halder}, \citenamefont {Gorman}, \citenamefont
  {Tapster}, \citenamefont {Rarity},\ and\ \citenamefont
  {Weinfurter}}]{kurtsiefer2002a}%
  \BibitemOpen
  \bibfield  {author} {\bibinfo {author} {\bibfnamefont {C.}~\bibnamefont
  {Kurtsiefer}}, \bibinfo {author} {\bibfnamefont {P.}~\bibnamefont {Zarda}},
  \bibinfo {author} {\bibfnamefont {M.}~\bibnamefont {Halder}}, \bibinfo
  {author} {\bibfnamefont {P.~M.}\ \bibnamefont {Gorman}}, \bibinfo {author}
  {\bibfnamefont {P.~R.}\ \bibnamefont {Tapster}}, \bibinfo {author}
  {\bibfnamefont {J.~G.}\ \bibnamefont {Rarity}}, \ and\ \bibinfo {author}
  {\bibfnamefont {H.}~\bibnamefont {Weinfurter}},\ }\href {\doibase
  10.1117/12.483036} {\bibfield  {journal} {\bibinfo  {journal} {Proc. SPIE}\
  }\textbf {\bibinfo {volume} {4917}},\ \bibinfo {pages} {25} (\bibinfo {year}
  {2002}{\natexlab{b}})}\BibitemShut {NoStop}%
\bibitem [{\citenamefont {Hughes}\ \emph {et~al.}(2002)\citenamefont {Hughes},
  \citenamefont {Nordholt}, \citenamefont {Derkacs},\ and\ \citenamefont
  {Peterson}}]{hughes2002}%
  \BibitemOpen
  \bibfield  {author} {\bibinfo {author} {\bibfnamefont {R.~J.}\ \bibnamefont
  {Hughes}}, \bibinfo {author} {\bibfnamefont {J.~E.}\ \bibnamefont
  {Nordholt}}, \bibinfo {author} {\bibfnamefont {D.}~\bibnamefont {Derkacs}}, \
  and\ \bibinfo {author} {\bibfnamefont {C.~G.}\ \bibnamefont {Peterson}},\
  }\href {\doibase 10.1088/1367-2630/4/1/343} {\bibfield  {journal} {\bibinfo
  {journal} {New J. Phys.}\ }\textbf {\bibinfo {volume} {4}},\ \bibinfo {pages}
  {43} (\bibinfo {year} {2002})}\BibitemShut {NoStop}%
\bibitem [{\citenamefont {Weier}\ \emph {et~al.}(2006)\citenamefont {Weier},
  \citenamefont {Schmitt-Manderbach}, \citenamefont {Regner}, \citenamefont
  {Kurtsiefer},\ and\ \citenamefont {Weinfurter}}]{weier2006}%
  \BibitemOpen
  \bibfield  {author} {\bibinfo {author} {\bibfnamefont {H.}~\bibnamefont
  {Weier}}, \bibinfo {author} {\bibfnamefont {T.}~\bibnamefont
  {Schmitt-Manderbach}}, \bibinfo {author} {\bibfnamefont {N.}~\bibnamefont
  {Regner}}, \bibinfo {author} {\bibfnamefont {C.}~\bibnamefont {Kurtsiefer}},
  \ and\ \bibinfo {author} {\bibfnamefont {H.}~\bibnamefont {Weinfurter}},\
  }\href {\doibase 10.1002/prop.200610322} {\bibfield  {journal} {\bibinfo
  {journal} {Fortschr. Phys.}\ }\textbf {\bibinfo {volume} {54}},\ \bibinfo
  {pages} {840} (\bibinfo {year} {2006})}\BibitemShut {NoStop}%
\bibitem [{\citenamefont {Ursin}\ \emph {et~al.}(2007)\citenamefont {Ursin},
  \citenamefont {Tiefenbacher}, \citenamefont {Schmitt-Manderbach},
  \citenamefont {Weier}, \citenamefont {Scheidl}, \citenamefont {Lindenthal},
  \citenamefont {Blauensteiner}, \citenamefont {Jennewein}, \citenamefont
  {Perdigues}, \citenamefont {Trojek}, \citenamefont {{\" O}mer}, \citenamefont
  {F{\" u}rst}, \citenamefont {Meyenburg}, \citenamefont {Rarity},
  \citenamefont {Sodnik}, \citenamefont {Barbieri}, \citenamefont
  {Weinfurter},\ and\ \citenamefont {Zeilinger}}]{ursin2007}%
  \BibitemOpen
  \bibfield  {author} {\bibinfo {author} {\bibfnamefont {R.}~\bibnamefont
  {Ursin}}, \bibinfo {author} {\bibfnamefont {F.}~\bibnamefont {Tiefenbacher}},
  \bibinfo {author} {\bibfnamefont {T.}~\bibnamefont {Schmitt-Manderbach}},
  \bibinfo {author} {\bibfnamefont {H.}~\bibnamefont {Weier}}, \bibinfo
  {author} {\bibfnamefont {T.}~\bibnamefont {Scheidl}}, \bibinfo {author}
  {\bibfnamefont {M.}~\bibnamefont {Lindenthal}}, \bibinfo {author}
  {\bibfnamefont {B.}~\bibnamefont {Blauensteiner}}, \bibinfo {author}
  {\bibfnamefont {T.}~\bibnamefont {Jennewein}}, \bibinfo {author}
  {\bibfnamefont {J.}~\bibnamefont {Perdigues}}, \bibinfo {author}
  {\bibfnamefont {P.}~\bibnamefont {Trojek}}, \bibinfo {author} {\bibfnamefont
  {B.}~\bibnamefont {{\" O}mer}}, \bibinfo {author} {\bibfnamefont
  {M.}~\bibnamefont {F{\" u}rst}}, \bibinfo {author} {\bibfnamefont
  {M.}~\bibnamefont {Meyenburg}}, \bibinfo {author} {\bibfnamefont
  {J.}~\bibnamefont {Rarity}}, \bibinfo {author} {\bibfnamefont
  {Z.}~\bibnamefont {Sodnik}}, \bibinfo {author} {\bibfnamefont
  {C.}~\bibnamefont {Barbieri}}, \bibinfo {author} {\bibfnamefont
  {H.}~\bibnamefont {Weinfurter}}, \ and\ \bibinfo {author} {\bibfnamefont
  {A.}~\bibnamefont {Zeilinger}},\ }\href {\doibase 10.1038/nphys629}
  {\bibfield  {journal} {\bibinfo  {journal} {Nat. Phys.}\ }\textbf {\bibinfo
  {volume} {3}},\ \bibinfo {pages} {481} (\bibinfo {year} {2007})}\BibitemShut
  {NoStop}%
\bibitem [{\citenamefont {Erven}\ \emph {et~al.}(2008)\citenamefont {Erven},
  \citenamefont {Couteau}, \citenamefont {Laflamme},\ and\ \citenamefont
  {G.Weihs}}]{erven2008}%
  \BibitemOpen
  \bibfield  {author} {\bibinfo {author} {\bibfnamefont {C.}~\bibnamefont
  {Erven}}, \bibinfo {author} {\bibfnamefont {C.}~\bibnamefont {Couteau}},
  \bibinfo {author} {\bibfnamefont {R.}~\bibnamefont {Laflamme}}, \ and\
  \bibinfo {author} {\bibnamefont {G.Weihs}},\ }\href {\doibase
  10.1364/oe.16.016840} {\bibfield  {journal} {\bibinfo  {journal} {Opt.
  Express}\ }\textbf {\bibinfo {volume} {16}},\ \bibinfo {pages} {16840}
  (\bibinfo {year} {2008})}\BibitemShut {NoStop}%
\bibitem [{\citenamefont {Peloso}\ \emph {et~al.}(2009)\citenamefont {Peloso},
  \citenamefont {Gerhardt}, \citenamefont {Ho}, \citenamefont {Lamas-Linares},\
  and\ \citenamefont {Kurtsiefer}}]{peloso2009}%
  \BibitemOpen
  \bibfield  {author} {\bibinfo {author} {\bibfnamefont {M.~P.}\ \bibnamefont
  {Peloso}}, \bibinfo {author} {\bibfnamefont {I.}~\bibnamefont {Gerhardt}},
  \bibinfo {author} {\bibfnamefont {C.}~\bibnamefont {Ho}}, \bibinfo {author}
  {\bibfnamefont {A.}~\bibnamefont {Lamas-Linares}}, \ and\ \bibinfo {author}
  {\bibfnamefont {C.}~\bibnamefont {Kurtsiefer}},\ }\href {\doibase
  10.1088/1367-2630/11/4/045007} {\bibfield  {journal} {\bibinfo  {journal}
  {New J. Phys.}\ }\textbf {\bibinfo {volume} {11}},\ \bibinfo {pages} {045007}
  (\bibinfo {year} {2009})}\BibitemShut {NoStop}%
\bibitem [{\citenamefont {Nauerth}\ \emph {et~al.}(2013)\citenamefont
  {Nauerth}, \citenamefont {Moll}, \citenamefont {Rau}, \citenamefont {Fuchs},
  \citenamefont {Horwath}, \citenamefont {Frick},\ and\ \citenamefont
  {Weinfurter}}]{nauerth2013}%
  \BibitemOpen
  \bibfield  {author} {\bibinfo {author} {\bibfnamefont {S.}~\bibnamefont
  {Nauerth}}, \bibinfo {author} {\bibfnamefont {F.}~\bibnamefont {Moll}},
  \bibinfo {author} {\bibfnamefont {M.}~\bibnamefont {Rau}}, \bibinfo {author}
  {\bibfnamefont {C.}~\bibnamefont {Fuchs}}, \bibinfo {author} {\bibfnamefont
  {J.}~\bibnamefont {Horwath}}, \bibinfo {author} {\bibfnamefont
  {S.}~\bibnamefont {Frick}}, \ and\ \bibinfo {author} {\bibfnamefont
  {H.}~\bibnamefont {Weinfurter}},\ }\href {\doibase 10.1038/NPHOTON.2013.46}
  {\bibfield  {journal} {\bibinfo  {journal} {Nat. Photonics}\ }\textbf
  {\bibinfo {volume} {7}},\ \bibinfo {pages} {382} (\bibinfo {year}
  {2013})}\BibitemShut {NoStop}%
\bibitem [{\citenamefont {Sajeed}\ \emph {et~al.}(2015)\citenamefont {Sajeed},
  \citenamefont {Radchenko}, \citenamefont {Kaiser}, \citenamefont {Bourgoin},
  \citenamefont {Pappa}, \citenamefont {Monat}, \citenamefont {Legr\'e},\ and\
  \citenamefont {Makarov}}]{sajeed2015}%
  \BibitemOpen
  \bibfield  {author} {\bibinfo {author} {\bibfnamefont {S.}~\bibnamefont
  {Sajeed}}, \bibinfo {author} {\bibfnamefont {I.}~\bibnamefont {Radchenko}},
  \bibinfo {author} {\bibfnamefont {S.}~\bibnamefont {Kaiser}}, \bibinfo
  {author} {\bibfnamefont {J.-P.}\ \bibnamefont {Bourgoin}}, \bibinfo {author}
  {\bibfnamefont {A.}~\bibnamefont {Pappa}}, \bibinfo {author} {\bibfnamefont
  {L.}~\bibnamefont {Monat}}, \bibinfo {author} {\bibfnamefont
  {M.}~\bibnamefont {Legr\'e}}, \ and\ \bibinfo {author} {\bibfnamefont
  {V.}~\bibnamefont {Makarov}},\ }\href {\doibase 10.1103/PhysRevA.91.032326}
  {\bibfield  {journal} {\bibinfo  {journal} {Phys. Rev. A}\ }\textbf {\bibinfo
  {volume} {91}},\ \bibinfo {pages} {032326} (\bibinfo {year}
  {2015})}\BibitemShut {NoStop}%
\bibitem [{\citenamefont {Jain}\ \emph {et~al.}(2014)\citenamefont {Jain},
  \citenamefont {Anisimova}, \citenamefont {Khan}, \citenamefont {Makarov},
  \citenamefont {Marquardt},\ and\ \citenamefont {Leuchs}}]{jain2014}%
  \BibitemOpen
  \bibfield  {author} {\bibinfo {author} {\bibfnamefont {N.}~\bibnamefont
  {Jain}}, \bibinfo {author} {\bibfnamefont {E.}~\bibnamefont {Anisimova}},
  \bibinfo {author} {\bibfnamefont {I.}~\bibnamefont {Khan}}, \bibinfo {author}
  {\bibfnamefont {V.}~\bibnamefont {Makarov}}, \bibinfo {author} {\bibfnamefont
  {C.}~\bibnamefont {Marquardt}}, \ and\ \bibinfo {author} {\bibfnamefont
  {G.}~\bibnamefont {Leuchs}},\ }\href {\doibase
  10.1088/1367-2630/16/12/123030} {\bibfield  {journal} {\bibinfo  {journal}
  {New J. Phys.}\ }\textbf {\bibinfo {volume} {16}},\ \bibinfo {pages} {123030}
  (\bibinfo {year} {2014})}\BibitemShut {NoStop}%
\bibitem [{\citenamefont {Jouguet}\ \emph {et~al.}(2013)\citenamefont
  {Jouguet}, \citenamefont {Kunz-Jacques},\ and\ \citenamefont
  {Diamanti}}]{jouguet2013}%
  \BibitemOpen
  \bibfield  {author} {\bibinfo {author} {\bibfnamefont {P.}~\bibnamefont
  {Jouguet}}, \bibinfo {author} {\bibfnamefont {S.}~\bibnamefont
  {Kunz-Jacques}}, \ and\ \bibinfo {author} {\bibfnamefont {E.}~\bibnamefont
  {Diamanti}},\ }\href {\doibase 10.1103/PhysRevA.87.062313} {\bibfield
  {journal} {\bibinfo  {journal} {Phys. Rev. A}\ }\textbf {\bibinfo {volume}
  {87}},\ \bibinfo {pages} {062313} (\bibinfo {year} {2013})}\BibitemShut
  {NoStop}%
\bibitem [{\citenamefont {Sun}\ \emph {et~al.}(2011)\citenamefont {Sun},
  \citenamefont {Jiang},\ and\ \citenamefont {Liang}}]{sun2011}%
  \BibitemOpen
  \bibfield  {author} {\bibinfo {author} {\bibfnamefont {S.-H.}\ \bibnamefont
  {Sun}}, \bibinfo {author} {\bibfnamefont {M.-S.}\ \bibnamefont {Jiang}}, \
  and\ \bibinfo {author} {\bibfnamefont {L.-M.}\ \bibnamefont {Liang}},\ }\href
  {\doibase 10.1103/PhysRevA.83.062331} {\bibfield  {journal} {\bibinfo
  {journal} {Phys. Rev. A}\ }\textbf {\bibinfo {volume} {83}},\ \bibinfo
  {pages} {062331} (\bibinfo {year} {2011})}\BibitemShut {NoStop}%
\bibitem [{\citenamefont {Lydersen}\ \emph {et~al.}(2010)\citenamefont
  {Lydersen}, \citenamefont {Wiechers}, \citenamefont {Wittmann}, \citenamefont
  {Elser}, \citenamefont {Skaar},\ and\ \citenamefont
  {Makarov}}]{lydersen2010a}%
  \BibitemOpen
  \bibfield  {author} {\bibinfo {author} {\bibfnamefont {L.}~\bibnamefont
  {Lydersen}}, \bibinfo {author} {\bibfnamefont {C.}~\bibnamefont {Wiechers}},
  \bibinfo {author} {\bibfnamefont {C.}~\bibnamefont {Wittmann}}, \bibinfo
  {author} {\bibfnamefont {D.}~\bibnamefont {Elser}}, \bibinfo {author}
  {\bibfnamefont {J.}~\bibnamefont {Skaar}}, \ and\ \bibinfo {author}
  {\bibfnamefont {V.}~\bibnamefont {Makarov}},\ }\href {\doibase
  10.1038/nphoton.2010.214} {\bibfield  {journal} {\bibinfo  {journal} {Nat.
  Photonics}\ }\textbf {\bibinfo {volume} {4}},\ \bibinfo {pages} {686}
  (\bibinfo {year} {2010})}\BibitemShut {NoStop}%
\bibitem [{\citenamefont {Zhao}\ \emph {et~al.}(2008)\citenamefont {Zhao},
  \citenamefont {Fung}, \citenamefont {Qi}, \citenamefont {Chen},\ and\
  \citenamefont {Lo}}]{zhao2008}%
  \BibitemOpen
  \bibfield  {author} {\bibinfo {author} {\bibfnamefont {Y.}~\bibnamefont
  {Zhao}}, \bibinfo {author} {\bibfnamefont {C.-H.~F.}\ \bibnamefont {Fung}},
  \bibinfo {author} {\bibfnamefont {B.}~\bibnamefont {Qi}}, \bibinfo {author}
  {\bibfnamefont {C.}~\bibnamefont {Chen}}, \ and\ \bibinfo {author}
  {\bibfnamefont {H.-K.}\ \bibnamefont {Lo}},\ }\href {\doibase
  10.1103/PhysRevA.78.042333} {\bibfield  {journal} {\bibinfo  {journal} {Phys.
  Rev. A}\ }\textbf {\bibinfo {volume} {78}},\ \bibinfo {eid} {042333}
  (\bibinfo {year} {2008})}\BibitemShut {NoStop}%
\bibitem [{\citenamefont {Qi}\ \emph {et~al.}(2007)\citenamefont {Qi},
  \citenamefont {Fung}, \citenamefont {Lo},\ and\ \citenamefont {Ma}}]{qi2007}%
  \BibitemOpen
  \bibfield  {author} {\bibinfo {author} {\bibfnamefont {B.}~\bibnamefont
  {Qi}}, \bibinfo {author} {\bibfnamefont {C.-H.~F.}\ \bibnamefont {Fung}},
  \bibinfo {author} {\bibfnamefont {H.-K.}\ \bibnamefont {Lo}}, \ and\ \bibinfo
  {author} {\bibfnamefont {X.}~\bibnamefont {Ma}},\ }\href@noop {} {\bibfield
  {journal} {\bibinfo  {journal} {Quant. Inf. Comp.}\ }\textbf {\bibinfo
  {volume} {7}},\ \bibinfo {pages} {73} (\bibinfo {year} {2007})}\BibitemShut
  {NoStop}%
\bibitem [{\citenamefont {Makarov}\ \emph {et~al.}(2006)\citenamefont
  {Makarov}, \citenamefont {Anisimov},\ and\ \citenamefont
  {Skaar}}]{makarov2006}%
  \BibitemOpen
  \bibfield  {author} {\bibinfo {author} {\bibfnamefont {V.}~\bibnamefont
  {Makarov}}, \bibinfo {author} {\bibfnamefont {A.}~\bibnamefont {Anisimov}}, \
  and\ \bibinfo {author} {\bibfnamefont {J.}~\bibnamefont {Skaar}},\ }\href
  {\doibase 10.1103/PhysRevA.74.022313} {\bibfield  {journal} {\bibinfo
  {journal} {Phys. Rev. A}\ }\textbf {\bibinfo {volume} {74}},\ \bibinfo
  {pages} {022313} (\bibinfo {year} {2006})},\ \bibinfo {note} {erratum ibid.
  \textbf{78}, 019905 (2008)}\BibitemShut {NoStop}%
\bibitem [{\citenamefont {Vakhitov}\ \emph {et~al.}(2001)\citenamefont
  {Vakhitov}, \citenamefont {Makarov},\ and\ \citenamefont
  {Hjelme}}]{vakhitov2001}%
  \BibitemOpen
  \bibfield  {author} {\bibinfo {author} {\bibfnamefont {A.}~\bibnamefont
  {Vakhitov}}, \bibinfo {author} {\bibfnamefont {V.}~\bibnamefont {Makarov}}, \
  and\ \bibinfo {author} {\bibfnamefont {D.~R.}\ \bibnamefont {Hjelme}},\
  }\href {\doibase 10.1080/09500340108240904} {\bibfield  {journal} {\bibinfo
  {journal} {J. Mod. Opt.}\ }\textbf {\bibinfo {volume} {48}},\ \bibinfo
  {pages} {2023} (\bibinfo {year} {2001})}\BibitemShut {NoStop}%
\bibitem [{\citenamefont {Gisin}\ \emph {et~al.}(2006)\citenamefont {Gisin},
  \citenamefont {Fasel}, \citenamefont {Kraus}, \citenamefont {Zbinden},\ and\
  \citenamefont {Ribordy}}]{gisin2006}%
  \BibitemOpen
  \bibfield  {author} {\bibinfo {author} {\bibfnamefont {N.}~\bibnamefont
  {Gisin}}, \bibinfo {author} {\bibfnamefont {S.}~\bibnamefont {Fasel}},
  \bibinfo {author} {\bibfnamefont {B.}~\bibnamefont {Kraus}}, \bibinfo
  {author} {\bibfnamefont {H.}~\bibnamefont {Zbinden}}, \ and\ \bibinfo
  {author} {\bibfnamefont {G.}~\bibnamefont {Ribordy}},\ }\href {\doibase
  10.1103/PhysRevA.73.022320} {\bibfield  {journal} {\bibinfo  {journal} {Phys.
  Rev. A}\ }\textbf {\bibinfo {volume} {73}},\ \bibinfo {pages} {022320}
  (\bibinfo {year} {2006})}\BibitemShut {NoStop}%
\bibitem [{\citenamefont {Lamas-Linares}\ and\ \citenamefont
  {Kurtsiefer}(2007)}]{lamas-linares2007}%
  \BibitemOpen
  \bibfield  {author} {\bibinfo {author} {\bibfnamefont {A.}~\bibnamefont
  {Lamas-Linares}}\ and\ \bibinfo {author} {\bibfnamefont {C.}~\bibnamefont
  {Kurtsiefer}},\ }\href {\doibase 10.1364/oe.15.009388} {\bibfield  {journal}
  {\bibinfo  {journal} {Opt. Express}\ }\textbf {\bibinfo {volume} {15}},\
  \bibinfo {pages} {9388} (\bibinfo {year} {2007})}\BibitemShut {NoStop}%
\bibitem [{\citenamefont {Weier}\ \emph {et~al.}(2011)\citenamefont {Weier},
  \citenamefont {Krauss}, \citenamefont {Rau}, \citenamefont {F{\"u}rst},
  \citenamefont {Nauerth},\ and\ \citenamefont {Weinfurter}}]{weier2011}%
  \BibitemOpen
  \bibfield  {author} {\bibinfo {author} {\bibfnamefont {H.}~\bibnamefont
  {Weier}}, \bibinfo {author} {\bibfnamefont {H.}~\bibnamefont {Krauss}},
  \bibinfo {author} {\bibfnamefont {M.}~\bibnamefont {Rau}}, \bibinfo {author}
  {\bibfnamefont {M.}~\bibnamefont {F{\"u}rst}}, \bibinfo {author}
  {\bibfnamefont {S.}~\bibnamefont {Nauerth}}, \ and\ \bibinfo {author}
  {\bibfnamefont {H.}~\bibnamefont {Weinfurter}},\ }\href {\doibase
  10.1088/1367-2630/13/7/073024} {\bibfield  {journal} {\bibinfo  {journal}
  {New J. Phys.}\ }\textbf {\bibinfo {volume} {13}},\ \bibinfo {pages} {073024}
  (\bibinfo {year} {2011})}\BibitemShut {NoStop}%
\bibitem [{\citenamefont {Fung}\ \emph {et~al.}(2009)\citenamefont {Fung},
  \citenamefont {Tamaki}, \citenamefont {Qi}, \citenamefont {Lo},\ and\
  \citenamefont {Ma}}]{fung2009}%
  \BibitemOpen
  \bibfield  {author} {\bibinfo {author} {\bibfnamefont {C.-H.~F.}\
  \bibnamefont {Fung}}, \bibinfo {author} {\bibfnamefont {K.}~\bibnamefont
  {Tamaki}}, \bibinfo {author} {\bibfnamefont {B.}~\bibnamefont {Qi}}, \bibinfo
  {author} {\bibfnamefont {H.-K.}\ \bibnamefont {Lo}}, \ and\ \bibinfo {author}
  {\bibfnamefont {X.}~\bibnamefont {Ma}},\ }\href@noop {} {\bibfield  {journal}
  {\bibinfo  {journal} {Quant. Inf. Comp.}\ }\textbf {\bibinfo {volume} {9}},\
  \bibinfo {pages} {131} (\bibinfo {year} {2009})}\BibitemShut {NoStop}%
\bibitem [{\citenamefont {Rau}\ \emph {et~al.}(2015)\citenamefont {Rau},
  \citenamefont {Vogl}, \citenamefont {Corrielli}, \citenamefont {Vest},
  \citenamefont {Fuchs}, \citenamefont {Nauerth},\ and\ \citenamefont
  {Weinfurter}}]{rau2015}%
  \BibitemOpen
  \bibfield  {author} {\bibinfo {author} {\bibfnamefont {M.}~\bibnamefont
  {Rau}}, \bibinfo {author} {\bibfnamefont {T.}~\bibnamefont {Vogl}}, \bibinfo
  {author} {\bibfnamefont {G.}~\bibnamefont {Corrielli}}, \bibinfo {author}
  {\bibfnamefont {G.}~\bibnamefont {Vest}}, \bibinfo {author} {\bibfnamefont
  {L.}~\bibnamefont {Fuchs}}, \bibinfo {author} {\bibfnamefont
  {S.}~\bibnamefont {Nauerth}}, \ and\ \bibinfo {author} {\bibfnamefont
  {H.}~\bibnamefont {Weinfurter}},\ }\href {\doibase
  10.1109/JSTQE.2014.2372008} {\bibfield  {journal} {\bibinfo  {journal} {IEEE
  J. Quantum Electron.}\ }\textbf {\bibinfo {volume} {21}},\ \bibinfo {pages}
  {6600905} (\bibinfo {year} {2015})}\BibitemShut {NoStop}%
\bibitem [{\citenamefont {Bourgoin}\ \emph {et~al.}()\citenamefont {Bourgoin},
  \citenamefont {Gigov}, \citenamefont {Higgins}, \citenamefont {Yan},
  \citenamefont {Meyer-Scott}, \citenamefont {Khandani}, \citenamefont {L{\"
  u}tkenhaus},\ and\ \citenamefont {Jennewein}}]{bourgoin2014}%
  \BibitemOpen
  \bibfield  {author} {\bibinfo {author} {\bibfnamefont {J.-P.}\ \bibnamefont
  {Bourgoin}}, \bibinfo {author} {\bibfnamefont {N.}~\bibnamefont {Gigov}},
  \bibinfo {author} {\bibfnamefont {B.~L.}\ \bibnamefont {Higgins}}, \bibinfo
  {author} {\bibfnamefont {Z.}~\bibnamefont {Yan}}, \bibinfo {author}
  {\bibfnamefont {E.}~\bibnamefont {Meyer-Scott}}, \bibinfo {author}
  {\bibfnamefont {A.}~\bibnamefont {Khandani}}, \bibinfo {author}
  {\bibfnamefont {N.}~\bibnamefont {L{\" u}tkenhaus}}, \ and\ \bibinfo {author}
  {\bibfnamefont {T.}~\bibnamefont {Jennewein}},\ }\href@noop {} {\ }\bibinfo
  {note} {({m}anuscript in preparation)}\BibitemShut {NoStop}%
\bibitem [{\citenamefont {Tyson}(2010)}]{tyson2010}%
  \BibitemOpen
  \bibfield  {author} {\bibinfo {author} {\bibfnamefont {R.}~\bibnamefont
  {Tyson}},\ }\href@noop {} {\emph {\bibinfo {title} {Principles of Adaptive
  Optics}}},\ \bibinfo {edition} {3rd}\ ed.\ (\bibinfo  {publisher} {{CRC
  Press}},\ \bibinfo {year} {2010})\BibitemShut {NoStop}%
\bibitem [{\citenamefont {Bienfang}\ \emph {et~al.}(2004)\citenamefont
  {Bienfang}, \citenamefont {Gross}, \citenamefont {Mink}, \citenamefont
  {Hershman}, \citenamefont {Nakassis}, \citenamefont {Tang}, \citenamefont
  {Lu}, \citenamefont {Su}, \citenamefont {Clark},\ and\ \citenamefont
  {Williams}}]{bienfang2004}%
  \BibitemOpen
  \bibfield  {author} {\bibinfo {author} {\bibfnamefont {J.~C.}\ \bibnamefont
  {Bienfang}}, \bibinfo {author} {\bibfnamefont {A.~J.}\ \bibnamefont {Gross}},
  \bibinfo {author} {\bibfnamefont {A.}~\bibnamefont {Mink}}, \bibinfo {author}
  {\bibfnamefont {B.~J.}\ \bibnamefont {Hershman}}, \bibinfo {author}
  {\bibfnamefont {A.}~\bibnamefont {Nakassis}}, \bibinfo {author}
  {\bibfnamefont {X.}~\bibnamefont {Tang}}, \bibinfo {author} {\bibfnamefont
  {R.}~\bibnamefont {Lu}}, \bibinfo {author} {\bibfnamefont {D.~H.}\
  \bibnamefont {Su}}, \bibinfo {author} {\bibfnamefont {C.~W.}\ \bibnamefont
  {Clark}}, \ and\ \bibinfo {author} {\bibfnamefont {C.~J.}\ \bibnamefont
  {Williams}},\ }\href {\doibase 10.1364/OPEX.12.002011} {\bibfield  {journal}
  {\bibinfo  {journal} {Opt. Express}\ }\textbf {\bibinfo {volume} {12}},\
  \bibinfo {pages} {2011} (\bibinfo {year} {2004})}\BibitemShut {NoStop}%
\bibitem [{\citenamefont {Takenaka}\ \emph {et~al.}(2012)\citenamefont
  {Takenaka}, \citenamefont {Toyoshima},\ and\ \citenamefont
  {Takayama}}]{takenaka2012}%
  \BibitemOpen
  \bibfield  {author} {\bibinfo {author} {\bibfnamefont {H.}~\bibnamefont
  {Takenaka}}, \bibinfo {author} {\bibfnamefont {M.}~\bibnamefont {Toyoshima}},
  \ and\ \bibinfo {author} {\bibfnamefont {Y.}~\bibnamefont {Takayama}},\
  }\href {\doibase 10.1364/OE.20.015301} {\bibfield  {journal} {\bibinfo
  {journal} {Opt. Express}\ }\textbf {\bibinfo {volume} {20}},\ \bibinfo
  {pages} {15301} (\bibinfo {year} {2012})}\BibitemShut {NoStop}%
\bibitem [{\citenamefont {Makarov}\ and\ \citenamefont
  {Hjelme}(2005)}]{makarov2005}%
  \BibitemOpen
  \bibfield  {author} {\bibinfo {author} {\bibfnamefont {V.}~\bibnamefont
  {Makarov}}\ and\ \bibinfo {author} {\bibfnamefont {D.~R.}\ \bibnamefont
  {Hjelme}},\ }\href {\doibase 10.1080/09500340410001730986} {\bibfield
  {journal} {\bibinfo  {journal} {J. Mod. Opt.}\ }\textbf {\bibinfo {volume}
  {52}},\ \bibinfo {pages} {691} (\bibinfo {year} {2005})}\BibitemShut
  {NoStop}%
\bibitem [{\citenamefont {Beaudry}\ \emph {et~al.}(2008)\citenamefont
  {Beaudry}, \citenamefont {Moroder},\ and\ \citenamefont
  {L\"{u}tkenhaus}}]{beaudry2008}%
  \BibitemOpen
  \bibfield  {author} {\bibinfo {author} {\bibfnamefont {N.~J.}\ \bibnamefont
  {Beaudry}}, \bibinfo {author} {\bibfnamefont {T.}~\bibnamefont {Moroder}}, \
  and\ \bibinfo {author} {\bibfnamefont {N.}~\bibnamefont {L\"{u}tkenhaus}},\
  }\href {\doibase 10.1103/PhysRevLett.101.093601} {\bibfield  {journal}
  {\bibinfo  {journal} {Phys. Rev. Lett.}\ }\textbf {\bibinfo {volume} {101}},\
  \bibinfo {eid} {093601} (\bibinfo {year} {2008})}\BibitemShut {NoStop}%
\bibitem [{\citenamefont {Tsurumaru}\ and\ \citenamefont
  {Tamaki}(2008)}]{tsurumaru2008}%
  \BibitemOpen
  \bibfield  {author} {\bibinfo {author} {\bibfnamefont {T.}~\bibnamefont
  {Tsurumaru}}\ and\ \bibinfo {author} {\bibfnamefont {K.}~\bibnamefont
  {Tamaki}},\ }\href {\doibase 10.1103/PhysRevA.78.032302} {\bibfield
  {journal} {\bibinfo  {journal} {Phys. Rev. A}\ }\textbf {\bibinfo {volume}
  {78}},\ \bibinfo {eid} {032302} (\bibinfo {year} {2008})}\BibitemShut
  {NoStop}%
\bibitem [{\citenamefont {Gittsovich}\ \emph {et~al.}(2014)\citenamefont
  {Gittsovich}, \citenamefont {Beaudry}, \citenamefont {Narasimhachar},
  \citenamefont {Alvarez}, \citenamefont {Moroder},\ and\ \citenamefont {L{\"
  u}tkenhaus}}]{gittsovich2014}%
  \BibitemOpen
  \bibfield  {author} {\bibinfo {author} {\bibfnamefont {O.}~\bibnamefont
  {Gittsovich}}, \bibinfo {author} {\bibfnamefont {N.~J.}\ \bibnamefont
  {Beaudry}}, \bibinfo {author} {\bibfnamefont {V.}~\bibnamefont
  {Narasimhachar}}, \bibinfo {author} {\bibfnamefont {R.~R.}\ \bibnamefont
  {Alvarez}}, \bibinfo {author} {\bibfnamefont {T.}~\bibnamefont {Moroder}}, \
  and\ \bibinfo {author} {\bibfnamefont {N.}~\bibnamefont {L{\" u}tkenhaus}},\
  }\href {\doibase 10.1103/PhysRevA.89.012325} {\bibfield  {journal} {\bibinfo
  {journal} {Phys. Rev. A}\ }\textbf {\bibinfo {volume} {89}},\ \bibinfo
  {pages} {012325} (\bibinfo {year} {2014})}\BibitemShut {NoStop}%
\bibitem [{\citenamefont {Marsili}\ \emph {et~al.}(2013)\citenamefont
  {Marsili}, \citenamefont {Verma}, \citenamefont {Stern}, \citenamefont
  {Harrington}, \citenamefont {Lita}, \citenamefont {Gerrits}, \citenamefont
  {Vayshenker}, \citenamefont {Baek}, \citenamefont {Shaw}, \citenamefont
  {Mirin},\ and\ \citenamefont {Nam}}]{marsili2013.NatPhotonics-7-210}%
  \BibitemOpen
  \bibfield  {author} {\bibinfo {author} {\bibfnamefont {F.}~\bibnamefont
  {Marsili}}, \bibinfo {author} {\bibfnamefont {V.~B.}\ \bibnamefont {Verma}},
  \bibinfo {author} {\bibfnamefont {J.~A.}\ \bibnamefont {Stern}}, \bibinfo
  {author} {\bibfnamefont {S.}~\bibnamefont {Harrington}}, \bibinfo {author}
  {\bibfnamefont {A.~E.}\ \bibnamefont {Lita}}, \bibinfo {author}
  {\bibfnamefont {T.}~\bibnamefont {Gerrits}}, \bibinfo {author} {\bibfnamefont
  {I.}~\bibnamefont {Vayshenker}}, \bibinfo {author} {\bibfnamefont
  {B.}~\bibnamefont {Baek}}, \bibinfo {author} {\bibfnamefont {M.~D.}\
  \bibnamefont {Shaw}}, \bibinfo {author} {\bibfnamefont {R.~P.}\ \bibnamefont
  {Mirin}}, \ and\ \bibinfo {author} {\bibfnamefont {S.~W.}\ \bibnamefont
  {Nam}},\ }\href {\doibase 10.1038/nphoton.2013.13} {\bibfield  {journal}
  {\bibinfo  {journal} {Nat. Photonics}\ }\textbf {\bibinfo {volume} {7}},\
  \bibinfo {pages} {210} (\bibinfo {year} {2013})}\BibitemShut {NoStop}%
\bibitem [{\citenamefont {Kerckhoffs}(1883)}]{kerckhoffs1883}%
  \BibitemOpen
  \bibfield  {author} {\bibinfo {author} {\bibfnamefont {A.}~\bibnamefont
  {Kerckhoffs}},\ }\href@noop {} {\bibfield  {journal} {\bibinfo  {journal} {J.
  des Sciences Militaires}\ }\textbf {\bibinfo {volume} {IX}},\ \bibinfo
  {pages} {5} (\bibinfo {year} {1883})}\BibitemShut {NoStop}%
\bibitem [{\citenamefont {Tomamichel}\ \emph {et~al.}(2012)\citenamefont
  {Tomamichel}, \citenamefont {Lim}, \citenamefont {Gisin},\ and\ \citenamefont
  {Renner}}]{tomamichel2012}%
  \BibitemOpen
  \bibfield  {author} {\bibinfo {author} {\bibfnamefont {M.}~\bibnamefont
  {Tomamichel}}, \bibinfo {author} {\bibfnamefont {C.~C.~W.}\ \bibnamefont
  {Lim}}, \bibinfo {author} {\bibfnamefont {N.}~\bibnamefont {Gisin}}, \ and\
  \bibinfo {author} {\bibfnamefont {R.}~\bibnamefont {Renner}},\ }\href
  {\doibase 10.1038/ncomms1631} {\bibfield  {journal} {\bibinfo  {journal}
  {Nat. Commun.}\ }\textbf {\bibinfo {volume} {3}},\ \bibinfo {pages} {634}
  (\bibinfo {year} {2012})}\BibitemShut {NoStop}%
\bibitem [{\citenamefont {Inamori}\ \emph {et~al.}(2007)\citenamefont
  {Inamori}, \citenamefont {L{\" u}tkenhaus},\ and\ \citenamefont
  {Mayers}}]{inamori2007}%
  \BibitemOpen
  \bibfield  {author} {\bibinfo {author} {\bibfnamefont {H.}~\bibnamefont
  {Inamori}}, \bibinfo {author} {\bibfnamefont {N.}~\bibnamefont {L{\"
  u}tkenhaus}}, \ and\ \bibinfo {author} {\bibfnamefont {D.}~\bibnamefont
  {Mayers}},\ }\href {\doibase 10.1140/epjd/e2007-00010-4} {\bibfield
  {journal} {\bibinfo  {journal} {Eur. Phys. J. D}\ }\textbf {\bibinfo {volume}
  {41}},\ \bibinfo {pages} {599} (\bibinfo {year} {2007})}\BibitemShut
  {NoStop}%
\bibitem [{\citenamefont {{da Silva}}\ \emph {et~al.}(2015)\citenamefont {{da
  Silva}}, \citenamefont {{do Amaral}}, \citenamefont {Xavier}, \citenamefont
  {Tempor{\~ a}o},\ and\ \citenamefont {{von der Weid}}}]{dasilva2015}%
  \BibitemOpen
  \bibfield  {author} {\bibinfo {author} {\bibfnamefont {T.~F.}\ \bibnamefont
  {{da Silva}}}, \bibinfo {author} {\bibfnamefont {G.~C.}\ \bibnamefont {{do
  Amaral}}}, \bibinfo {author} {\bibfnamefont {G.~B.}\ \bibnamefont {Xavier}},
  \bibinfo {author} {\bibfnamefont {G.~P.}\ \bibnamefont {Tempor{\~ a}o}}, \
  and\ \bibinfo {author} {\bibfnamefont {J.~P.}\ \bibnamefont {{von der
  Weid}}},\ }\href {\doibase 10.1109/JSTQE.2014.2361793} {\bibfield  {journal}
  {\bibinfo  {journal} {IEEE J. Sel. Top. Quantum Electron.}\ }\textbf
  {\bibinfo {volume} {21}},\ \bibinfo {pages} {1} (\bibinfo {year}
  {2015})}\BibitemShut {NoStop}%
\bibitem [{\citenamefont {Curty}\ \emph {et~al.}(2004)\citenamefont {Curty},
  \citenamefont {Lewenstein},\ and\ \citenamefont {L{\"
  u}tkenhaus}}]{curty2004a}%
  \BibitemOpen
  \bibfield  {author} {\bibinfo {author} {\bibfnamefont {M.}~\bibnamefont
  {Curty}}, \bibinfo {author} {\bibfnamefont {M.}~\bibnamefont {Lewenstein}}, \
  and\ \bibinfo {author} {\bibfnamefont {N.}~\bibnamefont {L{\" u}tkenhaus}},\
  }\href {\doibase 10.1103/PhysRevLett.92.217903} {\bibfield  {journal}
  {\bibinfo  {journal} {Phys. Rev. Lett.}\ }\textbf {\bibinfo {volume} {92}},\
  \bibinfo {pages} {217903} (\bibinfo {year} {2004})}\BibitemShut {NoStop}%
\bibitem [{\citenamefont {Hwang}(2003)}]{hwang2003}%
  \BibitemOpen
  \bibfield  {author} {\bibinfo {author} {\bibfnamefont {W.-Y.}\ \bibnamefont
  {Hwang}},\ }\href {\doibase 10.1103/PhysRevLett.91.057901} {\bibfield
  {journal} {\bibinfo  {journal} {Phys. Rev. Lett.}\ }\textbf {\bibinfo
  {volume} {91}},\ \bibinfo {pages} {057901} (\bibinfo {year}
  {2003})}\BibitemShut {NoStop}%
\end{thebibliography}
\end{document}